\documentclass[epj]{svjourmod}

\usepackage{graphicx,psfrag}
\usepackage{times}
\usepackage{color}
\usepackage{amsfonts,amssymb,stmaryrd,latexsym,amsmath}

\allowdisplaybreaks[1]

\newcommand{\beq}{\begin{equation}}
\newcommand{\eeq}{\end{equation}}
\newcommand{\nn}{\nonumber \\ }

\newcommand{\mpi}{M_\pi}

\newcommand{\qq}{\mathbf{Q}}
\newcommand{\pp}{\mathbf{p}}
\newcommand{\lb}{\mathbf{l}}
\newcommand{\rr}{\mathbf{r}}
\newcommand{\0}{\mathbf{0}}
\newcommand{\Order}{\mathcal{O}}

\begin{document}

\title{The multiple-scattering series in pion--deuteron scattering and  the nucleon--nucleon potential: perspectives from effective field theory}
\titlerunning{The multiple-scattering series in pion--deuteron scattering and  the nucleon--nucleon potential}

\author{V.~Baru\inst{1,2} \and E.~Epelbaum\inst{1} \and C.~Hanhart\inst{3,4}\and M.~Hoferichter\inst{5,6}
\and A.~E.~Kudryavtsev\inst{2}\and D.~R.~Phillips\inst{6}}
\authorrunning{V.~Baru {\it et al.}}

\institute{Institut f\"ur Theoretische Physik II, Ruhr-Universit\"at Bochum, D-44780 Bochum, Germany \and
Institute for Theoretical and Experimental Physics, B.\ Cheremushinskaya 25, 117218 Moscow, Russia \and
Institut f\"{u}r Kernphysik and J\"ulich Center for Hadron Physics, Forschungszentrum J\"{u}lich, D--52425 J\"{u}lich, Germany \and
Institute for Advanced Simulation, Forschungszentrum J\"{u}lich, D--52425 J\"{u}lich, Germany \and
Helmholtz-Institut f\"ur Strahlen- und Kernphysik and Bethe Center for Theoretical Physics, Universit\"at Bonn,  D--53115 Bonn, Germany \and
Institute of Nuclear and Particle Physics and Department of Physics and Astronomy, Ohio University, Athens, OH 45701, USA}

\date{
}

\abstract{
Important contributions to meson--nucleus scattering are produced by terms in
the multiple-scattering series, which is defined as the sum of all diagrams
where the meson scatters back and forth between a pair of static nucleons
before leaving the nucleus. In particular, the sum of this series is needed
for an accurate description of kaon--deuteron scattering, and appears as part
of the nucleon--nucleon potential. In this article we present some
effective-field-theory (EFT)-based insights into this series in the case of
two-nucleon systems.  In particular, we discuss the fact that, if meson--nucleon scattering 
is approximated by the scattering-length term, individual terms of the series
are divergent, and enhanced with respect to the straightforward expectation
from chiral perturbation theory ($\chi$PT). This apparently indicates the
presence of similarly enhanced counterterms. 
However, we show that when  the series is resummed the divergences
cancel, such that no additional information on short-range interactions is needed to obtain 
predictions for observables after resummation. 
We discuss the conditions under which this resummation is justified. 
We show that the same issues
arise in the $NN$ potential, where the resummed series produces poles
whose 
appearance indicates the breakdown scale of the $\chi$PT expansion
for that quantity. This demonstrates unequivocally that $\chi$PT  cannot be applied to compute $V(r)$
for distances smaller than $r \sim 1\,{\rm fm}$ at least in the theory
without explicit Delta(1232) degrees of freedom. We briefly discuss whether this bound can be
lowered if the Delta resonance is included in the EFT as an explicit degree of
freedom.}


\maketitle

\section{Introduction}

The multiple-scattering series (MSS) has played a prominent role in the study of meson--nucleus interactions.
Explicit expressions for the MSS have been known for a long time: first
derived in 1949 by Foldy in a different context~\cite{foldy}, it was
applied to $\pi d$ scattering as early as 1953 by Br\"uckner~\cite{brueck}.
The first diagrammatic representation was given in Ref.~\cite{kolkud} in 1972. 
More recently, the terms of the MSS have been shown to have a special status within an effective-field-theory treatment
 for $\pi$--nucleus scattering. The first EFT calculation of $\pi d$
 scattering, performed by Weinberg~\cite{weinberg}, classified the
 different contributions to the pion--deuteron scattering length,
 $a_{\pi d}$, according to their $\chi$PT order. Of the three three-body diagrams
 at leading order, by far the largest one is due to the second term in
 the $\pi d$ MSS, 
the so-called double-scattering term (see first diagram in~Fig.~\ref{fig:Feynman}). Weinberg's calculation has been refined in the twenty years since 
(see, e.g., Refs.~\cite{beane98,beane,recoil,disp,delta,Liebig,piD_PLB,longJOB}). In
Ref.~\cite{beane} it was observed that the triple-scattering term (i.e.~the third term
in the MSS) is significantly enhanced compared to what one would expect based on Weinberg's
original dimensional-analysis argument. In contrast to what was proposed in Ref.~\cite{beane}, it was shown in Ref.~\cite{Liebig} that the triple-scattering term is
enhanced by a factor of $\pi^2$ compared to its naive $\chi$PT order because of its special topology. The large contribution of the first diagram in~Fig.~\ref{fig:Feynman} to $a_{\pi d}$, together with the enhanced contribution of the second one, raises the question of whether all diagrams in the MSS are enhanced as compared to other $\chi$PT graphs in one way or another. 

\begin{sloppypar}
This question is of considerable contemporary import, as
data on mesonic atoms have, in recent years, become a prime source of experimental
information on strong meson--nucleon scattering lengths.
In particular, the pion--nucleon scattering
lengths were extracted from a combined analysis of pionic hydrogen~\cite{hydexp} and pionic
deuterium~\cite{deutexp} data with unprecedented accuracy in Refs.~\cite{piD_PLB,longJOB}.
Such an analysis calls for rigorous control over higher-order $\chi$PT corrections to pion--nucleus scattering, 
potentially  the most prominent of which are
the higher-order terms of the MSS.  Due to the smallness of the $\pi N$ scattering lengths, terms beyond triple scattering in the MSS give small contributions to $a_{\pi d}$. However, such a suppression does not show up
for $Kd$ scattering due to the relatively large $KN$ scattering lengths of the order of $1\,{\rm fm}$. In this case the 
non-perturbative resummation of all terms in the MSS is required~\cite{Ka01,Raha}\footnote{For a discussion 
of the role of recoil corrections potentially relevant for $Kd$ scattering, see Ref.~\cite{kdrec}.}.
The goal of this study is  to examine the consequences of enhancements of MSS terms for the $\chi$PT counting, and the grounds  for such a resummation.
\end{sloppypar}

Our main result is that care is required when expanding the MSS in a
diagrammatic fashion. In particular, 
a perturbative treatment of the series necessitates
the introduction of enhanced counterterms if well-defined expressions are
to be obtained. This is because the integrals appearing in the diagrammatic expansion
are individually divergent, starting from the quadruple-scattering term in the
series (third diagram in Fig.~\ref{fig:Feynman}). Nevertheless, we show
that, under particular circumstances, all those divergences cancel upon
resummation. Based on this observation we are 
also able to present a closed expression for the MSS in momentum space.

\begin{sloppypar}
The pertinent terms in the MSS also appear as sub-graphs within the $\chi$PT
contributions to the $NN$ potential. (It was pointed out long ago that the sum
of all two-particle irreducible $\pi NN \rightarrow \pi NN$ graphs itself
appears in $V_{NN}$~\cite{TR79,AM83,AB85,AP95,BK94B}.) Here we show that
``triangle graphs'' in the $NN$ potential are enhanced by factors of $\pi$
(not $\pi^2$) providing a special status to the diagrams of the MSS---and the
physics insights derived therefrom---in that problem, too.  We find that the
MSS contributions to $V_{NN}$ can be (partially) resummed. The effect of higher-order MSS
terms is minimal for $r \ge 1\,{\rm fm}$, but, for distances $r < 1\,{\rm
  fm}$, they produce unphysical poles in $V_{NN}(r)$. This leads us to suggest
that their appearance is associated with the breakdown of the $\chi$PT
expansion for the $NN$ potential at these distances.
\end{sloppypar}

The remainder of this paper is organized as follows. In Sect.~\ref{sec:mss} we
introduce the multiple-scattering series and define our conventions. In
Sect.~\ref{sec:pert} we perform a perturbative evaluation of the graphs in
this series and show that counterterms are needed in order to make sense of
the divergent momentum-space expressions which are encountered. In
Sect.~\ref{sec:regresumMSS} we provide a formal argument which vitiates the
need to consider these counterterms. We first regularize each term in the MSS,
then resum the series, and finally remove the regulator and obtain a finite
result. In Sect.~\ref{sec:inter} we discuss the limitations of this
procedure. In Sect.~\ref{sec:nn} we apply our insights from the meson--nucleus
case to the more complex case of the $NN$ potential. We offer our conclusions
in Sect.~\ref{sec:conc}.

\section{The multiple-scattering series}
\label{sec:mss}

\begin{figure}
\centering
\includegraphics[width=\linewidth]{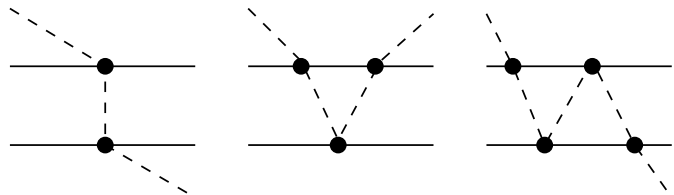}
\caption{Second, third, and fourth term in the MSS. Solid lines denote nucleons, dashed mesons, and
solid blobs interactions amongst them.}
\label{fig:Feynman}
\end{figure}

For the sake of simplicity, we start with isoscalar mesons scattering
off isoscalar nucleons. Then
the MSS for meson--nucleus scattering may be written 
as\footnote{The full MSS is given by $2 T$ because the pion can scatter from either nucleon to start any term in the MSS.}
\begin{equation}
2T =2(4\pi)^2\left(A^{(1)} + A^{(2)}+A^{(3)}+A^{(4)}+\cdots\right),
\label{eq:Aseries}
\end{equation}
where the superscript indicates the number of meson--nucleon interactions
appearing in the respective diagram, and a factor $2(4\pi)^2$ has been pulled out for 
convenience.
The second, third, and forth terms in the series are illustrated in
Fig.~\ref{fig:Feynman}. 

The $n$th term in the series shown in Fig.~\ref{fig:Feynman} has $n-2$ loops.
However, once nucleons are treated as static, the amplitude for a
zero-momentum pion scattering 
from the $NN$ system can be 
computed using the integral equation
\begin{align}
\label{eq:inteqn}
T({\bf p}',{\bf p})&=t_{\pi N}({\bf 0},{\bf 0}) (2 \pi)^3 \delta^{(3)}({\bf p}'-{\bf p}) \\
& + \int \frac{d^3 p''}{(2 \pi)^3} t_{\pi N}({\bf p}-{\bf p}'',{\bf 0})\frac{1}{({\bf p}-{\bf p}'')^2} T({\bf p}',{\bf p}''),\nonumber
\end{align}
where ${\bf p}$ (${\bf p}'$) is the relative momentum between the 
incoming (outgoing) $NN$ pair
and $t_{\pi N}({\bf q}',{\bf q})$ describes a $\pi N$ interaction with final (initial) pion momentum ${\bf q}'$ (${\bf q}$).
The contribution to the $\pi d$ scattering length stemming from the multiple-scattering series, Eq.~\eqref{eq:inteqn}, 
then reads\footnote{Throughout this work we use nuclear-physics
  conventions for the scattering lengths. Note that this differs from the
  conventions employed in previous studies of $\pi d$ scattering~\cite{beane,Liebig,piD_PLB,longJOB}.}
\beq
\label{conv}
a_{\pi d} 
=-\frac{1}{16\pi^4} \int d^3p\,d^3p'\Psi^\dagger(\pp')T({\bf p}',{\bf p})\Psi(\pp),
\eeq
where $\Psi(\pp)$ denotes the deuteron wave function normalized as   
\beq
\int d^3p\,\Psi^\dagger(\pp)\Psi(\pp)=1\label{deut_norm}.
\eeq

If only low-momentum components of the nuclear system are being probed, we are
encouraged to take
\begin{equation}
t_{\pi N}({\bf 0},{\bf 0})=t_{\pi N}({\bf p}-{\bf p}'',{\bf 0})=-4 \pi a,
\label{eq:scattlength}
\end{equation}
where $a$ is the $\pi N$ scattering length. (Note that we neglected the kinematical prefactors  suppressed as $\mpi/m_N$ with $m_N$
being the nucleon mass.)
In  $\chi$PT the absolute value of the isovector $\pi N$ scattering length is
\begin{equation}
\label{piN_scatt}
a=\frac{\mpi}{8\pi F_\pi^2},
\end{equation}
a result that will be needed below for the power
counting, as an example of a natural meson--nucleon scattering length,
 even though we mainly consider isoscalar meson--nucleon scattering\footnote{We do not consider the isoscalar 
 $\pi N$ scattering length, since it is unnaturally small both due to its chiral suppression and due to significant numerical cancellations.}.

\section{Perturbative evaluation}
\label{sec:pert}

If $t_{\pi N}$ has the form~\eqref{eq:scattlength}, 
the first two terms of the MSS may be written as 
\begin{align}
A^{(1)}(\qq)&=-2 \pi^2 a \delta^{(3)}(\qq),\notag\\
A^{(2)}(\qq)&=\frac{a^2}{Q^2}, 
\end{align}
while terms $A^{(m)}(\qq)$ starting with $m=3$ have the form 
\begin{align}
\label{eq:Am}
A^{(m)}(\qq)  &= \frac{(-4\pi a)^m}{(4\pi)^2}\int \frac{d^3l_1}{(2\pi)^3}\cdots \frac{d^3l_{m-2}}{(2\pi)^3}\ \frac{1}{\lb_1^2}\\
&\hspace{-15pt}\times \left[ \frac{1}{(\lb_1-\lb_2)^2}\frac{1}{(\lb_{2}-\lb_{3})^2}\cdots\frac{1}{(\lb_{m-2}-\qq)^2}\right].\notag
\end{align}
Here, $\qq={\bf p}'-{\bf p}$ denotes the three-momentum transfer between the incoming and 
outgoing nucleons and $Q=|\qq|$. The third term, for example, reads
\begin{equation}
\label{a3expr}
A^{(3)}(\qq)=-\frac{\pi}{2}\frac{a^3}{Q}.
\end{equation}
Upon
Fourier transforming we find
\begin{equation}
A^{(2)}(r)=\frac{a^2}{4 \pi r}, \quad A^{(3)}(r)=-\frac{a^3}{4 \pi r^2}.
\end{equation}
These are the first two cases of the well-known form for the $m$th term
\begin{equation}
A^{(m)}(r)=-\frac{a}{4 \pi} \left(-\frac{a}{r}\right)^{m-1}.
\label{eq:Amr}
\end{equation}

\begin{sloppypar}
However, a problem occurs when trying to derive the forth term from the momentum-space integral~\eqref{eq:Am}
\beq
\label{a4expr}
A^{(4)}(\qq)=\int\frac{d^3l_1}{(2\pi)^3}\frac{d^3l_2}{(2\pi)^3}
\frac{(4\pi)^2 a^4}{\lb_1^2(\lb_1-\lb_2)^2(\lb_2-\qq)^2}.
\eeq
Since the integral itself is dimensionless, and yet depends on only one
dimensionful parameter, $Q$, it should come as no surprise that the
result of the integral is independent of that dimensionful parameter. Indeed, one finds,
after introducing the variable $z=l_2/Q$,
\begin{equation} 
\label{A4}
A^{(4)}(\qq)=a^4\left(1+\int_1^\infty \frac{dz}{z}\right).
\end{equation} 
The Fourier transform of the first, constant, term is a delta-function in configuration
space, which does not match the expected $r$-space expression $a^4/4\pi r^3$. Moreover, the remaining integral in Eq.~\eqref{A4} is divergent. The easiest way to regularize it is to introduce a finite upper
limit in the integration, ${z_{\rm max}=\mu/Q}$, which, equivalently, translates to a finite cutoff
in $l_2$.  In this way, through the regularization of the integral, the
momentum transfer $Q$ again appears in the expression. 
\end{sloppypar}

\begin{sloppypar}
The necessity for regularization of the integral implies, from an EFT perspective, the need to introduce a $\pi d$ counterterm in order to parameterize the short-distance physics which is affecting the result. Its operator structure is a constant in momentum space. This introduces a free
parameter: the strength of the finite piece of the counterterm. To
simplify the notation we may absorb the constant in the expression
given above into this free parameter. We then find
\begin{equation} 
A^{(4)}(\qq)=-a^4\log\frac{Q}{\mu}+\frac{f_0(\mu)}{32\pi^2},
\label{eq:a4mom}
\end{equation} 
where $f_0$ is a $\mu$-dependent parameter. Note that the factor $32\pi^2$ appears here due to our conventions for the definition
of the individual terms in the MSS according to Eqs.~\eqref{eq:Aseries} and~\eqref{eq:Am}. 
\end{sloppypar}

The first term in Eq.~\eqref{eq:a4mom}, relative to $A^{(2)}$, is then of order
\beq
a^2Q^2\sim\frac{\mpi^4}{64\pi^2F_\pi^4}=4\pi^2\left(\frac{\mpi}{4\pi F_\pi}\right)^4,
\eeq
where we have counted $Q\sim\mpi$. The suppression of $A^{(4)}$ relative to $A^{(2)}$ is therefore $4\pi^2$ less than the naive-dimensional-analysis (NDA) estimate for such a two-loop effect, a manifestation of the $\pi^2$ enhancement alluded to above.  

The size of $f_0$ is determined by the coefficient of the logarithm in Eq.~\eqref{eq:a4mom}. We thus have
\begin{align}
f_0 \sim 32\pi^2a^4 = \frac{\mpi^4}{128\pi^2 F_\pi^8}= \frac{2\pi^2}{F_\pi^4}
\left(\frac{\mpi}{4\pi F_\pi}\right)^4.
\label{f0}
\end{align} 
$f_0$ then, like the other term in Eq.~\eqref{eq:a4mom}, is $2\pi^2$ larger than 
$\chi$PT power counting would have suggested for a counterterm that renormalizes 
a two-loop diagram. The enhanced (with respect to their $\chi$PT estimate) size 
of the MSS terms therefore potentially drives the existence of larger-than-expected short-distance effects. 

\begin{sloppypar}
As an example of what happens with higher terms in the MSS, we
briefly mention the corresponding result for $A^{(5)}$. Straightforward
evaluation gives
\begin{equation}
A^{(5)}(\qq)=a^5\frac{\pi}{4}\left(Q-2\int dl\right). 
\end{equation}
The first term in the brackets, which is finite, can
be mapped onto the $1/r^4$ term of the MSS via a 
properly regularized Fourier transform, while the
second (divergent) term generates, in principle, another free  parameter. Note, however, that this free parameter, 
linear in $\mu$, can be absorbed into $f_0$. This pattern continues, with all terms of even order in the MSS 
apparently requiring new operator structures to make them finite, while the odd-order terms have finite pieces 
which map properly onto their coordinate-space expressions, and whose divergent parts can be absorbed into the 
counterterms generated at the preceding order in the MSS. 
\end{sloppypar}

There is thus an apparent problem, since the results obtained above
 would imply  that the MSS comes with an infinite set of free parameters, 
and cannot be regarded as predictive. Moreover, these counterterms are 
larger than one would expect
based on $\chi$PT counting for the $\pi d$ problem, because the MSS terms
themselves are larger than implied by naive application of $\chi$PT.

\section{Resumming the MSS}

\label{sec:regresumMSS}

In this section we show how these problems can be resolved, the central finding being
that the UV divergences of the MSS cancel exactly once the series is resummed---for all values of $Q$. 
A direct consequence of this result is that no enhanced counterterms are required 
in the MSS. The  limitations on the validity of this result  will be addressed in more detail in the next section.

In order to proceed we first regularize each term in the MSS and then sum the entire series. 
This allows us to recover the standard result for each term in the series 
once the regulator is removed and the resummed result re-expanded in powers of $a$.

Our starting point is Eq.~\eqref{eq:Am}. Using the fact that 
\beq
\frac{1}{p^2}=\frac{1}{4\pi}\int d^3 r \frac{\exp(i \pp\cdot\rr)}{r},
\eeq
and applying this to each individual propagator in Eq.~\eqref{eq:Am},
after  integration over all three momenta and angles
 one obtains, for the $m$th term in the MSS ($m \geq 2$)
\begin{equation}
A^{(m)}(\qq) = (-a)^m \int \frac{dr}{r^{m-3}}\frac{\sin Qr}{Qr}.
\label{eq:AmQ}
\end{equation}
This expression exhibits the following types of divergences: first, 
for $Q = 0$ and $m = 4$ it becomes singular in the infrared
(IR). In practice, this singularity gets tamed automatically once the convolution
with the nuclear wave functions is included (cf.~Eq.~\eqref{conv}). Second, 
starting from $m = 4$, this expression is singular in the
ultraviolet (UV), $r \to 0$, as mentioned above. We will show in the 
following that  all UV divergences cancel once the MSS is resummed. 
Finally, this resummation produces
a new type of singularity that shows up for $a < 0$, and
will be discussed in~Sect.~\ref{sec:inter}. In what follows we choose
a regularization technique which makes the divergences manifest, so that the argument can be
presented in a straightforward and clean way.

In particular, we now regularize~\eqref{eq:AmQ} by IR and UV cutoffs $R$ and $r_0$, respectively.  
Beginning with the evaluation of the Fourier transform at $Q=0$ we see that the regularized version of $A^{(m)}(\0)$ is
\begin{equation}
A^{(m)}(\0) = (-a)^m \int_{r_0}^R \frac{dr}{r^{m-3}},
\label{aatQ0}
\end{equation}
which gives, for $m=4$,
\begin{equation}
\label{eq1}
A^{(4)}(\0) ={a^4} \log\frac{R}{r_0}.
\end{equation}
Thus this integral shows a logarithmic divergence as $r_0\to 0$, in full correspondence to what was stated above (cf.~the $\log{\mu}$ 
in Eq.~\eqref{eq:a4mom}). On the other hand, for $m>4$ we obtain
\begin{equation}
\label{eq2}
A^{(m)}(\0) = {(-a)^m}\int^R_{r_0} \frac{dr}{r^{m-3}}= \frac{(-a)^m}{m-4}\bigg(\frac{1}{r_0^{m-4}}-\frac{1}{R^{m-4}}\bigg).
\end{equation}
The sum of all orders beyond order 4 in $A^{(m)}(\0)$ is then
\begin{equation}
\sum_{m>4}A^{(m)}(\0) = {a^4}\left\{\log\frac{R+a}{r_0+a}+\log\frac{r_0}{R}\right\}.
\label{eq:sumabove4}
\end{equation}
The last, singular, term in this expression cancels exactly with $A^{(4)}(\0)$.
Thus, we find that the final result at $Q=0$ is UV finite provided that
regularization is carried out and the full series then resummed. After
resummation the limit $r_0\to0$ is finite so that the regulator can be
formally removed, if $a>0$.  For $a<0$ this is not possible, since one would
hit the branch cut of the logarithm in the first term of
Eq.~\eqref{eq:sumabove4}.  However, as long
as the scattering length is natural,
$|a| \lesssim 1/\Lambda$, the regulator $r_0$ can at least be pushed outside the regime
of validity of the theory. In any case there is no need
for the inclusion of enhanced (compared to their $\chi$PT estimate) counterterms to remove the UV divergences
in the resummed expression.   
The physical implications of this procedure will be
discussed in Sect.~\ref{sec:inter}.

\begin{sloppypar}
The divergence structure of the case $Q \neq 0$ can be
reduced to that discussed in the previous paragraph. Let
\begin{equation}
A^{(m)}(\qq) = A^{(m)}(\0)+\delta A^{(m)}(\qq). 
\end{equation}
The terms $A^{(m)}(\0)$ were already dealt with above.
In addition, $\delta A^{(m)}(\qq)$ is UV finite for $m=4$ and $m=5$,
however, starting from $m=6$ also these terms diverge. Let
\begin{align}\nonumber
\delta \hat A(\qq)  &=   \sum_{m>5} \delta A^{(m)}(\qq) \\ 
&=  
\sum_{m>5}{(-a)^m}\int_{r_0}^R\frac{dr}{r^{m-3}}\left(\frac{\sin Qr}{Qr}-1\right).
\end{align}
In order to proceed we now expand $\sin Qr$ in a power series around
$Q=0$
and study each term individually. We get
\begin{equation}
\delta \hat A(\qq)  =
\sum_{n=1}^{\infty}\sum_{m>5}{(-a)^m}\frac{(-1)^nQ^{2n}}{(2n{+}1)!}\int_{r_0}^Rdrr^{3-m+2n}.
\end{equation}
This expression becomes UV singular for $m\geq 2n+4$. For those terms we may write
\begin{align} \nonumber
\delta \hat A(\qq)_{\rm sing}  &=  
\sum_{n=1}^{\infty}\sum_{l=1}^\infty{(-a)^{l+2n+3}}\frac{(-1)^nQ^{2n}}{(2n{+}1)!}\int_{r_0}^R\frac{dr}{r^l} \\
 &=  \left(\frac{\sin Qa}{Qa}-1\right) 
  \sum_{l=1}^\infty{(-a)^{l+3}}\int_{r_0}^R\frac{dr}{r^l}.
\end{align}
The last sum is the same one we encountered in summing the terms $A^{(m)}(0)$ (see Eq.~\eqref{aatQ0}) from $m=4$ to $\infty$.  We may
therefore follow the same steps applied in that case to obtain
\begin{equation}
\delta \hat A(\qq)_{\rm sing}  =
\left(\frac{\sin Qa}{Qa}-1\right)
{a^4}\log\frac{R+a}{r_0+a},
\end{equation}
which again has a smooth limit for $r_0\to 0$---as long as $a>0$. Thus we have demonstrated
that, for all values of $Q$, the UV divergences of the MSS cancel exactly once
the series is resummed.  Thus, apparently no counter\-terms are required and
the MSS is predictive. No additional information on short-distance physics is
needed in order for it to render a sensible prediction, as long as we
regularize it, and then resum. The UV regulator can then be removed.
\end{sloppypar}

Thus far we have been able to sum the singular terms and construct an explicit
momentum-space expression for them after summation. This is not, however, the
full sum of the MSS, which also
includes those terms whose Fourier transform is well-defined without any need to
introduce a regulator. In order to derive the full result we solve the integral equation~\eqref{eq:inteqn} 
and hence obtain an expression for the full sum of the MSS
in coordinate and momentum space.

We can, under quite general conditions,
solve Eq.~\eqref{eq:inteqn} by a function
\begin{equation}
2 T({\bf p}',{\bf p})=2 (4\pi)^2 A(\qq).
\label{eq:ansatz}
\end{equation}
The function $A$ is most easily computed by taking the inverse Fourier
transform of Eq.~\eqref{eq:inteqn} and applying the convolution theorem. This
produces~\cite{Ka01}
\begin{equation}
A(r)=-\frac{a}{4\pi} - \frac{a}{r} A(r),
\end{equation}
such that, as was shown long ago~\cite{brueck,kolkud},
\begin{equation}
\label{eq:fullsum}
A(r)=-\frac{a r}{4\pi(r+a)}=-\frac{a}{4 \pi} + \frac{a^2}{4\pi r}\sum_{n=0}^{\infty}\left(-\frac{a}{r}\right)^n,
\end{equation}
and thus each term in the MSS contributes 
one order in a geometric series in $a/r$. 

Clearly, the final expression given in Eq.~\eqref{eq:fullsum} is a very
efficient and useful representation of the MSS.  It is not singular as $r
\rightarrow 0$ --- at least as long as $a>0$ --- although the individual terms
of the sum are increasingly singular. The evaluation of the expectation value of the full sum with nuclear wave
functions is straightforward---again, as long as $a > 0$.

The easiest way to obtain the resummed MSS in momentum space is
to perform a Fourier transform of the resummed $r$-space expression~\eqref{eq:fullsum}. This yields
\begin{equation}
{A(\qq)} = -\frac{(2\pi)^3}{4\pi}a \delta^{(3)}(\qq)+\frac{a^2}{Q^2} - \frac{a^3}{Q} f(aQ),
\label{eq:AQfull}
\end{equation}
with
\begin{align}
\nonumber
f(y) &= \int_0^\infty dx\frac{\sin x}{x+y} \\
&=  \mbox{Ci}(y)\sin y +\frac12 \cos y(\pi-2\mbox{Si}(y)),
\label{eq:fdef}
\end{align}
where we used the following definitions of the sine and cosine
integral functions
\begin{equation}
\mbox{Ci}(y) = -\int_y^\infty dt \frac{\cos t}{t},\quad 
\mbox{Si}(y) = \int_0^y dt\frac{\sin t}{t}.
\end{equation}

In this context it is important to note that the function Ci$(y)$, appearing
in Eq.~\eqref{eq:fdef}, has a branch point
at $y=0$ --- it acquires an unphysical cut from $aQ=0$ to
$aQ=-\infty$ which enters the physical region for negative values of $a$.
This cut appears to be the momentum space analog of the unphysical pole at ${r=-a}$
of Eq.~\eqref{eq:fullsum}, but it should not be a concern for $a > 0$.

Moreover, 
for small positive $y$,  we may expand $f(y)$ in powers
 of $y$, and so obtain
\begin{equation}
f(aQ) = \frac{\pi}{2}+(\gamma-1+\log aQ)aQ-\frac{\pi (aQ)^2}{4}+ \cdots,
\label{eq:expandf}
\end{equation}
where $\gamma$ denotes the Euler--Mascheroni constant. Insertion of this expansion into 
Eq.~\eqref{eq:AQfull} produces a power series in $a$ in which
all non-analytic terms match with
what we found in Sect.~\ref{sec:pert}. Note, however, that in the full, finite expression
the logarithmic divergence of Eq.~\eqref{eq:a4mom} becomes effectively regularized at the scale $1/a$.
To better understand the structure and the coefficients
of the term $\propto a^4$ in Eq.~\eqref{eq:AQfull}, one may calculate the quadruple-scattering term explicitly starting from the expression~\eqref{eq:AmQ}
\begin{align}
\nonumber
A^{(4)}(\qq) &= {a^4} \int_{Q r_0}^{\infty} 
{d x} \frac{\sin x}{x^2}\\
&=-a^4\, (\gamma-1+\log Q r_0).
\label{eq:AmQ4}
\end{align}
This explains how the constant term $\gamma-1$ appears in Eq.~\eqref{eq:expandf} and demonstrates that
the UV regulator $r_0$ is effectively replaced by $a$ due to the
resummation procedure described above. 

The pole/cut in coordinate/momentum space that appears for $a<0$ for the MSS
of meson--nucleus scattering emerges since we focused on isoscalar
interactions. Under certain conditions, a different isospin structure can make
the pole disappear. To illustrate this point, we consider $\pi^-d$ scattering
with the $\pi^-p\to \pi^0n$ and $\pi^0n\to\pi^0n$ channels switched off. In the
isospin limit, this can be described with isoscalar and isovector $\pi N$
scattering lengths $a^+$ and $a^-$. The (coordinate space) result for the
resummed MSS then reads 
\beq
A(r)=-\frac{a^+}{4\pi}+\frac{(a^+)^2-(a^-)^2}{4\pi(r^2-(a^+)^2+(a^-)^2)}(r-a^+),
\eeq 
which for $a^-\to0$ reduces to Eq.~\eqref{eq:fullsum}. Provided that
$|a^-|>|a^+|$, the pole disappears and the result for the MSS is well-defined
everywhere. Similarly, a pole in the full MSS for $\pi^-d$ scattering, with
the $\pi^-p\to \pi^0n$ and $\pi^0n\to\pi^0n$ channels included, would only
appear if $\pi N$ interactions were not predominantly of isovector nature.

For this reason, the discussion of the case $a<0$ in which a pole appears in $A(r)$
might appear quite academic. However, as we will show in Sect.~\ref{sec:nn},
this kind of pole does appear when the sum of the $\pi NN$ MSS which contributes to the $NN$ potential
is computed.

\section{Interpretation}
\label{sec:inter}

The previous section suggests that the MSS has a valid expansion 
 if $0 < aQ < 1$, so that the expression~\eqref{eq:expandf} can be employed and  the cut in $f(y)$ 
does not enter the physical $(Q>0)$ region. In coordinate space these conditions 
correspond to $r > |a|$ and $a > 0$. In case of $a<0$, however, a pole (cut)
appears in the resummed coordinate (momentum) space expression.
This issue will be addressed at the end of this section.

\begin{sloppypar}
In general, the fact that a resummation gives a well-defined answer does not
necessarily mean that it gives the correct answer.  This was recently stressed
in Ref.~\cite{Epelbaum:2009sd}. In particular, one might be concerned that
the sensitivity to short-distance physics in the MSS necessitated by the
perturbative treatment with a point-like $\pi N$ interaction
indicates that the regularized, resummed result of Sect.~\ref{sec:regresumMSS} obtained in the limit 
of $r_0\to 0$ is not correct and thus strongly reduces the predictive
power of our theoretical approach. 
\end{sloppypar} 

In general, the $\pi N$ interaction has a certain range,  and 
the $t$-matrix is not momentum independent.
 Here we therefore introduce a regularized $\pi N$
interaction, which is non-pointlike at scale $\Lambda$, and discuss the
conditions under which the results of the previous section are accurate.
In this way the scale  $\Lambda$ is completely equivalent to  the regulator 
$r_0$ introduced in the previous section, however, it has a physical  interpretation of the scale that 
limits the validity of the theory.   
Thus the difference between the results with a finite $\Lambda$ and those in
the limit $\Lambda\to \infty$   provides a measure of the consistency of the scheme. 
This difference, if the scheme is self-consistent, should be less than the contribution of 
the leading contact operator. 
For example, for pion--deuteron scattering investigated within $\chi$PT the first contact term 
contributes at  $\Order(p^2)$ (or parametrically ${m_{\pi}^2}/({ f_{\pi}^4}{ \Lambda^2})$) 
and its size was estimated to be around $5\,\%$ relative to the leading 
double-scattering diagram~\cite{Liebig,longJOB}.  

We now write
\begin{equation}
\label{def_reg}
t_{\pi N}({\bf p},{\bf 0})=-4 \pi a \hat g\left(\frac{|{\bf p}|}{\Lambda}\right)
\end{equation}
with $\hat g(x) \rightarrow 1$ as $x\to 0$. 
The amplitude for scattering of the zero-momentum pion from the $NN$ system still takes the form~\eqref{eq:ansatz}, but now
\begin{equation}
A(r)=-\frac{a}{4 \pi} - \frac{a}{r} g(r) A(r)
\label{Arform}
\end{equation}
with 
\begin{equation}
g(r)=4 \pi r \int \frac{d^3 k}{(2 \pi)^3} e^{-i {\bf k} \cdot {\bf r}} \frac{1}{{\bf k}^2}  \hat g^2\left(\frac{|{\bf k}|}{\Lambda}\right).
\end{equation}
Thus, in the presence of the regulator $g$,
\begin{equation}
A(r)=-\frac{ar}{4\pi(r + a g(r))}.
\label{eq:regMSS}
\end{equation}
If, for illustrative purposes, we take
\begin{equation}
\hat g(x)=\frac{1}{\sqrt{1+x^2}},
\label{eq:dipole}
\end{equation}
then 
\begin{equation}
g(r)=1-e^{-\Lambda r},
\label{eq:fr}
\end{equation}
and so $g(r) \rightarrow 1$ as $r \rightarrow \infty$, but $g(r) \rightarrow
\Lambda r$ as $r \rightarrow 0$. While the details of this result are
specific to the form~\eqref{eq:dipole}, the disappearance of regulator
effects in the infrared and the appearance of additional powers of $r$ in the
ultraviolet is a general feature.

If $A(r)$ is evaluated perturbatively based on the expansion of Eq.~\eqref{eq:regMSS}, 
we have (cf.~Eq.~\eqref{eq:fullsum})
\begin{equation}
A^{(m)}(r)=-\frac{a}{4 \pi} \left(-\frac{ag(r)}{r}\right)^{m-1}
\label{eq:Amrpert}
\end{equation}
with $m\ge 1$.
Observe that introducing  the form factor $g$ with the cutoff $\Lambda$ 
leads to results completely equivalent to those obtained using a sharp cutoff that were discussed in detail in the previous section. In particular, the individual terms 
in the MSS  again appear to be enhanced relative to their $\chi$PT estimates: for example, the momentum-space form
of the quadruple-scattering term  exhibits the 
behavior $A^{(4)}(\qq) \sim -a^4 \log Q/\Lambda$  discussed in Sect.~\ref{sec:pert}. 
And, once again,  the resummed result~\eqref{eq:regMSS} is much less
$\Lambda$-dependent than the individual terms in the MSS. The limit  $\Lambda\to \infty$ exists there due to the cancellations of the  UV-divergent  terms
derived in the previous section. But, the form \eqref{eq:Amrpert} makes explicit that it is
the behavior $g(r) \rightarrow \Lambda r$ as $r \rightarrow 0$ which guarantees that none of the terms in the MSS diverge as $r \rightarrow 0$. 
It is crucial to observe that
this regularization only prevails if $\Lambda$ is kept finite, as an effective scale representing 
the range of validity of the theory, when
Eq.~\eqref{eq:regMSS} is expanded in powers of $a$ and Fourier transformed to momentum space. 
Although, of course, if one is interested in the long-distance ($r > |a|$) form of each contribution 
to the MSS in $r$-space, the limit $\Lambda \rightarrow \infty$ can be safely taken in each of the terms in 
Eq.~\eqref{eq:Amrpert}.

\begin{sloppypar}
In physical terms we anticipate
working in perturbative meson--nucleon systems where $|a| \sim 1/\Lambda$, and so we can resum
the series in $a/r$ to infinite order with $\Lambda$ being large but finite. 
This yields a well-defined mathematical procedure.   
In the resummed expression (see Eq.~\eqref{eq:regMSS}) 
the limit $\Lambda \rightarrow \infty$ exists. 
However, the question is whether it can be safely taken.  
In other words, how does the physics of the finite $\Lambda$  affect the EFT result for the MSS? The difference between the $\Lambda
 \rightarrow \infty$ and finite-$\Lambda$ results for $A(r)$ is
\begin{equation}
\Delta A \equiv A(r)-A_{\Lambda \rightarrow \infty}(r)=-\frac{a^2 r (1 - g(r))}{4\pi(r+a)(r+ag(r))}.
\label{eq:Delta}
\end{equation}
For $\pi d$ scattering within $\chi$PT,
the convolution of Eq.~\eqref{eq:Delta} with the (pionful) deuteron wave functions
obtained in chiral EFT~\cite{NNLO} 
results in  an effect of less than $3\,\%$ to the pion--deuteron scattering length 
(with $\Lambda\sim M_{\rho}\sim 800 $ MeV). 
This is fully in line with the  estimate of the contact operator at $\Order(p^2)$. 
Thus, taking the limit $\Lambda\to \infty$ is justified in the resummed expression~\eqref{eq:regMSS}, 
and the results of the previous section are correct. No  enhanced counterterms 
are therefore required in the case of the MSS with a natural scattering length. 
\end{sloppypar}

\begin{figure}
\centering
\includegraphics[width=\linewidth]{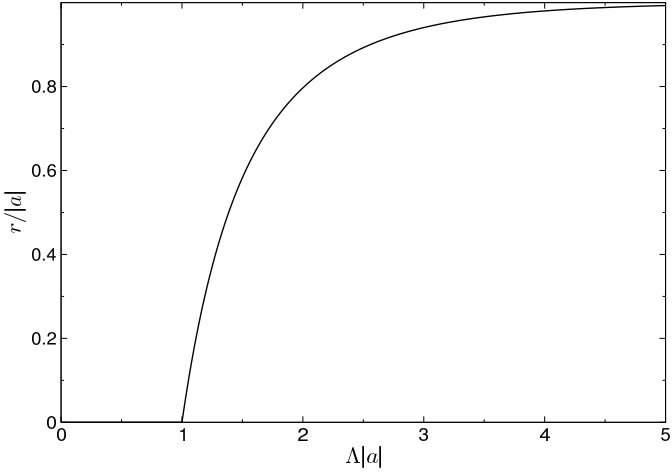}
\caption{Position of the pole at $r=-a g(r)$ for $a<0$. The pole is at $r=0$ if $\Lambda |a| \leq 1$.}
\label{fig:pole}
\end{figure}

\begin{sloppypar}
We now focus our attention on the pole at ${r=-a}$, relevant 
in the case $a<0$. In Eq.~\eqref{eq:regMSS} the analog of this pole occurs as the solution of
\begin{equation}
r=-ag(r),
\label{pole}
\end{equation}
which can be found in  an analytic form for the form factor chosen in~Eq.~\eqref{eq:fr}.
In this case, Eq.~\eqref{pole} can be rewritten in the form 
\beq
\Lambda (r-|a|)\, e^{\Lambda (r-|a|)}=-\Lambda |a| e^{-\Lambda |a|},
\eeq
whose solution is 
\beq
r=\frac{1}{\Lambda} \left(\Lambda |a| +W\big(-\Lambda |a| e^{-\Lambda |a|}\big)\right),
\label{rsol}
\eeq
where $W(z)$ is the Lambert $W$-function defined  via
$z= W e^W$ for any complex number $z$.  
It has a branch-cut discontinuity in the complex plane $z$ running from $-\infty$ to $-1/e$. 
In particular,  the branch point $z=-1/e$ corresponds to $\Lambda |a|=1$ in Eq.~\eqref{rsol}. As a consequence, 
for $\Lambda |a| \le 1 $ we have  $W(-\Lambda |a| e^{-\Lambda |a|}) =-\Lambda |a|$ and the pole is at the origin.    
In contrast, for  $\Lambda |a| > 1 $ the pole moves away from $0$ and is located within the interval $r\in (0, |a|]$, the upper limit 
being reached in the situation $\Lambda |a| \gg 1 $. The corresponding trajectory of the pole as a function of $\Lambda|a|$ is depicted in Fig.~\ref{fig:pole}.
\end{sloppypar}
 
For a natural value of the meson--nucleon scattering length the pole only arises due to short-distance 
($r \sim 1/\Lambda$) parts of the MSS diagrams, and these parts of the diagrams cannot be regarded as 
a reliable prediction of the EFT. Or, phrased differently, we can assume that
for values of $r$ in the applicability range of the theory the resummed MSS
delivers reliable results as long as $a$ is natural. 

An   unnaturally large scattering length $a$, i.e.\ $|a| \gg 1/\Lambda$,
 indicates
the presence of a shallow meson--nucleon bound or virtual state. In this case, once $r
\sim a$ it is not appropriate to write $t=-4 \pi a$, even if $r \gg
1/\Lambda$. 
Instead we must at least resum unitarity corrections to the meson--nucleon
amplitude, which should remove the pole in the case of $a <0$ or shift it towards the origin. 
It remains to be seen in each particular case whether it suffices 
to keep only the leading unitarity term ($\propto ik$) in the amplitude, or if a 
non-perturbative treatment of finite-range corrections is needed as well.

As a possible example of this situation one may consider $Kd$ scattering where at least one of
the $KN$ scattering lengths is of  the order of $1\,{\rm fm}$ and negative, see, e.g.,~Ref.~\cite{Raha}. 
In this case it appears useful to employ a non-relativistic effective field theory~\cite{Raha,kdrec}
which operates with the threshold parameters ($KN$ scattering lengths) and reproduces the 
result of the $Kd$ MSS~\cite{Ka01}. The usefulness of the theory is based on the separation 
of two distinct scales.  While the $NN$ interaction is mediated by one-pion exchange at large distances,
the $KN$ interaction is governed by the two-pion exchange, which justifies the treatment of $KN$ 
interactions as point-like. The range of validity of the approach is thus $\Lambda\sim 300$ MeV. 
In spite of the large scattering length, for such a small $\Lambda$ the product $\Lambda|a|$ is still close to $1$
so that the pole might still be near the origin\footnote{Note that in the real world the $KN$ scattering lengths 
are strongly absorptive so that the pole cannot be on the real axis.}. Therefore we do not expect the counterterm
to be enhanced in this case either.
Meanwhile, keeping the unitarity corrections at distances $r \sim a$ appears 
necessary.

\section{Remarks on the multiple-scattering series in the nucleon--nucleon potential}
\label{sec:nn}

We now turn our attention to the MSS in the nucleon--nucleon potential. After
a pion is emitted from one nucleon it can propagate in the $NN$ system via the
full MSS, before being reabsorbed on the other nucleon. Therefore, also in the
nucleon--nucleon potential, diagrams enhanced compared to their $\chi$PT order
in a similar manner to that discussed above appear. In this section we discuss
the consequences of this aspect of the MSS for chiral EFT computations of the
$NN$ potential.  Here, we leave aside the issues associated with
non-perturbative renormalization of that potential. The interested reader may
consult
Refs.~\cite{Epelbaum:2009sd,Lepage:1997cs,Birse:2005um,Nogga:2005hy,PavonValderrama:2005wv,Epelbaum:2006pt}
and references therein for a sample of different views on this issue.  In this
work we are interested in two particular questions regarding the
meson-exchange diagrams which generate the long-distance (van-der-Waals in
the chiral limit) part of the potential: are the MSS diagrams enhanced? If so,
what does that imply for the scale at which a perturbative expansion of the
long-distance potential breaks down?

A key difference between
the multiple-scattering terms in $\pi$--nucleus scattering and the
nucleon--nucleon potential is that the meson propagator $1/(\lb_i-\lb_j)^2$ as
it appears, e.g., in Eq.~\eqref{a4expr} is to be replaced by
${1/((\lb_i-\lb_j)^2+\mpi^2)}$. The expression
for the corresponding one-loop term is then proportional to~\cite{Ordonez:1995rz,KBW97}
\begin{equation}
\frac{1}{2Q}\arctan\frac{Q}{2\mu},
\label{nnscatvspiAscat}
\end{equation}
with $\mu=\mpi$. In the kinematics for pion--nucleus
scattering we need to choose $\mu=0$, since here
the energy transfer and the meson mass cancel exactly in the pion propagator. Once this limit is taken an additional factor of $\pi/2$ appears, and Eq.~\eqref{nnscatvspiAscat} reduces
to Eq.~\eqref{a3expr}. Thus, the enhancement of this graph in the $NN$
scattering potential is not the $\pi^2$ we found for $\pi$--nucleus
scattering, but we do still have enhancement by a factor of $\pi$ over the NDA estimate of this graph.  

This enhancement is phenomenologically important.  It is well known that the
strongest contribution to the two-pion-exchange potential up to N$^2$LO
emerges from the subleading triangle diagram. While nominally subleading, the
corresponding central isoscalar potential appears to be an order of magnitude
stronger than all the other two-pion-exchange contributions. This unnaturally
large contribution can be traced back to the aforementioned triangle graphs'
enhancement by one power of $\pi$, together with the numerically large value
of the low-energy constant (LEC) $c_3$, which parameterizes the subleading
$\pi\pi NN$ vertex and is largely saturated by the $\Delta$ isobar~\cite{Bernard:1996gq}. 
These observations provide a strong motivation to take
a closer look at higher-order terms in the MSS beyond the triangle
diagram. Although one expects that potentials generated by the exchange of a large
number of pions are exponentially suppressed at distances $r \sim \mpi^{-1}$,
one should keep in mind their singular, van-der-Waals-like behavior at shorter
distances.  We will see below how these competing features influence the
convergence of the chiral expansion for this particular set of diagrams.
 
\begin{figure}
\centering
\includegraphics[width=0.7\linewidth]{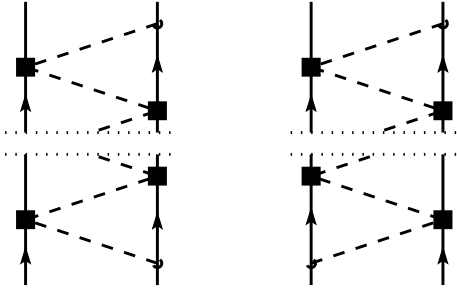}
\caption{Time-ordered MSS-diagram contribution to the nucleon--nucleon
  potential. Solid dots (filled rectangles) refer to the leading-order vertices
  from the effective Lagrangian proportional to $g_A$ ($c_i$).  }
\label{fig:Feynman_time_ordered}
\end{figure}

\begin{sloppypar}
To be specific, we consider time-ordered two-nucleon diagrams in the MSS
as shown in Fig.~\ref{fig:Feynman_time_ordered}. 
In this exploratory study, we restrict ourselves to
the isoscalar part of the subleading $\pi \pi NN$ vertex 
\begin{equation}
\langle \lb_1 ,  i | H | \lb_2 ,  j \rangle = 
\frac{\delta_{ij}}{F_\pi^2 \sqrt{\omega_1 \omega_2}} \left( 2 c_1 \mpi^2 + c_3 \lb_1
  \cdot \lb_2 \right),
\end{equation}
where $\lb_{1,2}$ denote the pion momenta, $i$ and $j$ are the
pion isospin quantum numbers and  $\omega_{1,2} \equiv \sqrt{\lb_{1,2}^2 +
  \mpi^2}$ are the free pion energies.  The potential corresponding to the
left diagram in Fig.~\ref{fig:Feynman_time_ordered}, where an even number of pion exchanges occurs, is given by 
\begin{align}
V^{n\pi}(\mathbf{q}) &= - \frac{3 g_A^2}{8 F_\pi^{2n}} \,
\int \frac{d^3 l_1}{(2 \pi)^3} \ldots \frac{d^3 l_n}{(2 \pi)^3} \, (2 \pi)^3
\nn &\times \delta^{(3)} (\lb_1 + \lb_2 + \ldots + \lb_n - {\bf q} )  
\frac{\boldsymbol{\sigma}_2
  \cdot \lb_n \, \boldsymbol{\sigma}_2 \cdot \lb_1}{\omega_1^2 \omega_2^2 \ldots
\omega_n^2} \nn
&\times  
( 2 c_1 \mpi^2 - c_3 \lb_1 \cdot \lb_2) \, ( 2 c_1 \mpi^2 - c_3 \lb_2
\cdot \lb_3) \nn
 &\times  \ldots \,  ( 2 c_1 \mpi^2 - c_3 \lb_{n-1} \cdot \lb_n)
\, + \, 1 \leftrightarrow 2.
\end{align}
Clearly, the integrals entering this expression are UV divergent. The
divergences, however, are absorbed into $4N$ contact operators, and so do not affect $r$-space expressions for the finite-range part of the
potential that we are discussing here. The $r$-space
representation of the potential can be obtained straightforwardly, leading to 
\begin{align}
V^{n\pi} (\rr ) &= \frac{3 g_A^2}{4 F_\pi^{2n}} \boldsymbol{\nabla}_1 \cdot \boldsymbol{\nabla}_n 
\, ( 2 c_1 \mpi^2 + c_3 \boldsymbol{\nabla}_1 \cdot \boldsymbol{\nabla}_2) \nn
&\times  ( 2 c_1 \mpi^2 + c_3 \boldsymbol{\nabla}_2 \cdot \boldsymbol{\nabla}_3) \ldots \nn
&\times  ( 2 c_1 \mpi^2 + c_3 \boldsymbol{\nabla}_{n-1} \cdot \boldsymbol{\nabla}_n)\\
&\times  U(r_1)  \, U(r_2) \, \ldots \, U (r_n) \,  \Big|_{r_1 = r_2 = \ldots
  = r_n = r}, 
\nonumber
\end{align}
with
\beq
U (r) = \frac{1}{4 \pi r} \, e^{-\mpi r}
\eeq
being the usual Yukawa function. After evaluating the derivatives, one ends up
with the isoscalar central potential 
\begin{align}
\label{VCentIsosc}
V^{n\pi} (\rr ) &= \frac{3 g_A^2}{4 (4 \pi F_\pi^{2})^n} \, \frac{e^{-nx}}{r^{3n}}
\Bigg[ \sum_{m=0}^{n-2} \sum_{l=0}^{m} y^n_{ml}(2 c_1 x^2)^{n-m-1} \nn
&\times  c_3^m (1 + x)^{2(m +1-l)} (2 + 2x + x^2)^l \nn
&+ c_3^{n-1} \left( (2 + 2x + x^2)^n + 2(1+x)^n \right)\Bigg].
\end{align}
Here we introduced a dimensionless variable $x \equiv \mpi r$ and 
combinatorial coefficients $y^n_{ml}$ whose explicit values can be derived
straightforwardly.  In a completely similar way, one finds that the second
diagram in Fig.~\ref{fig:Feynman_time_ordered}, where an odd number of pion exchanges takes place, gives rise to the isovector tensor and spin-spin
potential 
\begin{align}
V^{n\pi} (\rr) &= \frac{g_A^2}{4 (4 \pi F_\pi^{2})^n} \, \boldsymbol{\tau}_1 \cdot \boldsymbol{\tau}_2 \, \frac{e^{-nx}}{r^{3n}} \, \Bigg( \boldsymbol{\sigma}_1 \cdot \hat{\rr} \, 
\boldsymbol{\sigma}_2 \cdot \hat{\rr}  \nn
&\times \Bigg[\sum_{m=0}^{n-2} \sum_{l=0}^{m} y^n_{ml} (2 c_1 x^2)^{n-m-1}
c_3^m (1 + x)^{2(m +1-l)} \nn
&\times  (2 + 2x + x^2)^l + c_3^{n-1} \left( (2 + 2x + x^2)^n \right. \nn 
&+  \left. (1+x)^n \right) \Bigg] - \boldsymbol{\sigma}_1 \cdot \boldsymbol{\sigma}_2 \, 
c_3^{n-1}\, (1+x)^n \Bigg),
\end{align}
where $n=2k+1, \, k\in \mathbb{N}$. 

As expected, based on the discussion at the start of this section, each extra loop in the MSS generates a power of
$1/(4 \pi F_\pi^2)$, rather than the $1/(4 \pi F_\pi)^2$ that is usually assumed in
$\chi$PT. This is the way the ``enhancement'' of MSS diagrams plays out in the $NN$ potential. 
These contributions to the  potential, $V^{n\pi} (\rr)$,  take a particularly simple form if either $c_1$
or $c_3$ is set to zero. In particular, the central isoscalar potential in
Eq.~\eqref{VCentIsosc} reads in these two limits 
\begin{align}
\label{simple_case}
V_{c_1}^{n \pi} (\rr ) &= \frac{3 g_A^2}{8 (2 \pi F_\pi^{2})^n} \, \frac{e^{-nx}}{r^{3n}}
(c_1 x^2)^{n-1} (1+x)^2 , \\
\mbox{\hspace{-0.7cm}}
V_{c_3}^{n \pi} (\rr ) &= \frac{3 g_A^2 c_3^{n-1} }{4 (4 \pi F_\pi^{2})^n} \, \frac{e^{-nx}}{r^{3n}}
\left[(2 + 2x + x^2)^n + 2 (1+x)^n \right].
\nonumber
\end{align}
Resumming the resulting geometric series leads to the following closed-form
expressions for the potentials
\begin{align}
\label{V3V1}
V_{c_1} (\rr) &= \frac{3 g_A^2 c_1 \mpi^2}{32 \pi^2 F_\pi^4} \, \frac{e^{-2x}}{r^4}
(1+x)^2 \, \frac{1}{1 - \frac{c_1^2 \mpi^4}{4 \pi^2 F_\pi^4} \,
  \frac{e^{-2x}}{r^2}}, \nn
V_{c_3} (\rr) &= \frac{3 g_A^2 c_3 }{64 \pi^2 F_\pi^4} \,
\frac{e^{-2x}}{r^6} \left[ 
\frac{2(1+x)^2}{1 - \frac{c_3^2}{16 \pi^2 F_\pi^4} \,
  \frac{e^{-2x}}{r^6}(1+x)^2 } \right. \nn
&+  \left.\frac{(2+2x+x^2)^2}{1 - \frac{c_3^2}{16 \pi^2 F_\pi^4}  \,
   \frac{e^{-2x}}{r^6} (2+2x+x^2)^2 } \right].
\end{align}
Both $V_{c_3}$ and $V_{c_1}$ feature poles at finite values of $r >
0$ similar to what we observed for the MSS of meson--nucleus scattering
for $a<0$ --- c.f.~Eq.~\eqref{eq:fullsum} --- only here the
appearance of the poles is independent of the sign of the scattering
parameters $c_i$. 
These unphysical poles are non-perturbative phenomena resulting from the partial resummation
of the amplitude.  
\end{sloppypar}

\begin{figure}
\centering
\includegraphics[width=\linewidth]{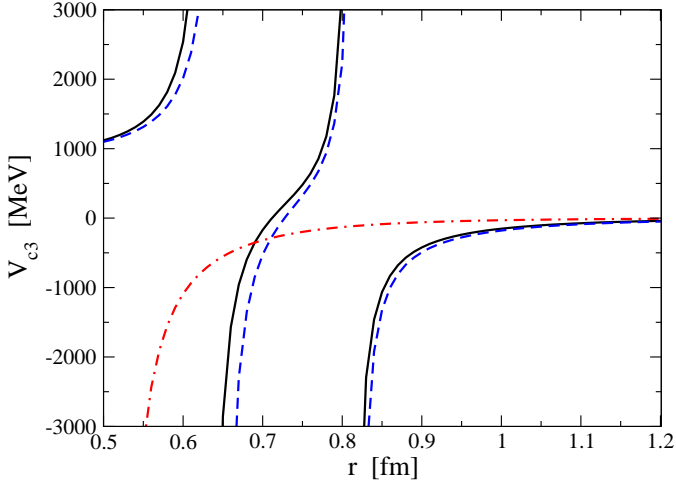}
\caption{Resummed potential $V_{c_3}$ from Eq.~\eqref{V3V1}. The solid (dashed) line shows the
  potential for $c_3=-3.87$ GeV$^{-1}$ and $\mpi=138$ MeV ($\mpi=138/4$
  MeV), while the dashed-dotted line depicts the potential for  $c_3=-1$ GeV$^{-1}$ and
  $\mpi=138$ MeV. }
\label{fig:vc3}
\end{figure}

\begin{sloppypar}
One may view the location of the poles as a measure of  the
breakdown scale of the chiral expansion for the considered class of
diagrams.
 It is comforting to see that the pole in  $V_{c_1}$ is located at a
rather short distance, namely 
\beq
r \sim \frac{|c_1 | \, \mpi^2}{2\pi F_\pi^2} \sim 0.05\,{\rm fm},
\eeq
and is shifted to the origin in the chiral limit. On the other hand, the
pole positions in $V_{c_3}$ are not protected by powers of $\mpi$ and
can be estimated by
\beq
\label{pole_extim}
r \sim \mathcal{O} \left( \left( \frac{| c_3 |}{\pi F_\pi^2} \right)^{1/3}
\right) \sim  \mathcal{O} \left( 1\,{\rm fm} \right),
\eeq
using the value $c_3 = -3.87$ GeV$^{-1}$ from the $\Order(Q^2)$ fit to
$\pi N$ threshold coefficients of Ref.~\cite{Krebs:2007rh}. Numerically, the poles appearing in the 
two terms of $V_{c_3}$ in Eq.~\eqref{V3V1} 
are found to be located at 
\beq
r \simeq 0.63\,{\rm fm},\quad \quad
r \simeq 0.81\,{\rm fm}, 
\eeq   
see Fig.~\ref{fig:vc3}. 
As shown in this figure, the behavior of the resummed potential at
short distances and the pole positions only weakly depend on the values of the
pion mass in the case of $V_{c_3}$. It is somewhat surprising that the chiral
expansion for the pion-exchange potential breaks down at the relatively large
distance of $\sim 0.8\,{\rm fm}$. This behavior is, to a large extent,
caused by the already discussed enhancement of the diagrams in the MSS, where 
loops generate inverse powers of $4 \pi F_\pi^2$ rather than $(4 \pi F_\pi
)^2$, see Eq.~\eqref{pole_extim}. An additional enhancement occurs due to
the large numerical value of the LEC $c_3$. As shown in~\cite{Krebs:2007rh}, this LEC
takes a much more natural value of the order $c_3 \sim -1\,{\rm GeV}^{-1}$ once the
$\Delta$-isobar is explicitly taken into account. Therefore, one might expect
that the unphysical poles in the potential are shifted closer to the origin in
the $\Delta$-full approach, see Fig.~\ref{fig:vc3}. For example, setting 
$c_3 = -1\,{\rm GeV}^{-1}$ the poles are shifted to 
\beq
r \simeq 0.41\,{\rm fm},\quad \quad
r \simeq 0.52\,{\rm fm}. 
\eeq   
The above arguments therefore suggest that the breakdown
scale for the chiral expansion of the pion-exchange potential 
is in the range $r \sim 0.5 \ldots 0.8\,{\rm fm}$. This estimate agrees well
with the findings of various recent studies, see e.g.~\cite{Valderrama:2009ei,Birse:2010jr,Ipson:2010ah,Valderrama:2011mv}.   
In this context it should be stressed, however, that it would be insufficient
to include the $\Delta$ as a static field, for then its inclusion would do
nothing but to restore the original strength of $c_3$ --- see discussion in
Refs.~\cite{KGW98,Krebs:2007rh}. Thus, only a $\Delta$ with retained recoils could
help,
which makes sense only if also the nucleons are treated as non-static. The
corresponding
calculations are very involved and go beyond the scope of this paper.
\end{sloppypar}

\begin{figure}
\centering
\includegraphics[width=\linewidth]{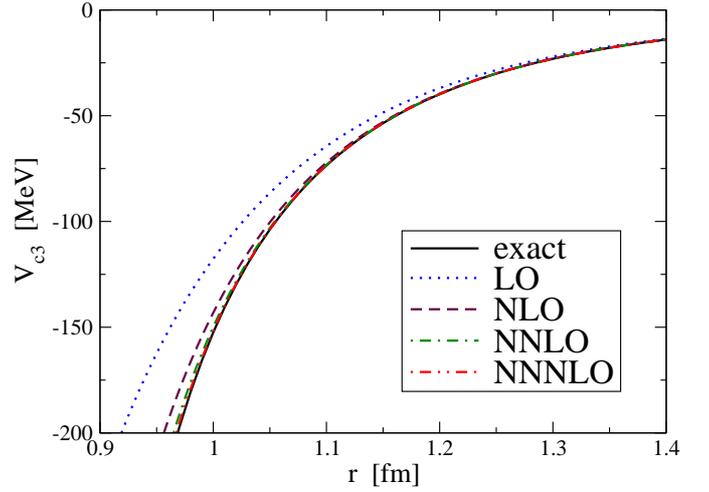}
\caption{Convergence of the MSS for $V_{c_3}$.}
\label{fig:conv}
\end{figure}

Given the rather large value of the breakdown scale $r \sim 0.8\,{\rm fm}$, one might
worry about the convergence of the chiral expansion for the potential 
at distances of the order $1 \ldots 2\,{\rm fm}$. Fortunately,
the convergence of the MSS appears to be rather fast, see Fig.~\ref{fig:conv}. 
In particular, one observes that the potential is already very well described
by the subleading term in the MSS.  Clearly, the reason for this fast
convergence is due to the exponential falloff of the potential at distances
large compared to the exchanged mass. 

\begin{sloppypar}
It must be stressed that, while the results obtained here for the resummed
potential provide qualitative insights, they are by no means a complete $\chi$PT calculation.
In addition to the omitted baryon recoils,
 we only picked out time-ordered graphs that give
rise to the MSS. In contrast to near-threshold\footnote{More precisely, the
  argument refers to the case when the momenta of external pions and nucleons
are much smaller than $\mpi$.} pion--nucleus scattering, where
time-ordered graphs involving two or more pions in the intermediate states can
be represented by contact operators~\cite{Beane:2000fi}, there is, strictly speaking, no
justification for neglecting such diagrams in the $NN$ potential. On the other
hand, the neglected time-ordered graphs in most cases are suppressed by powers
of $\pi$ compared to the one in the MSS. 
This is because the maximal enhancement by powers of $\pi$
requires that the individual pion exchanges factorize out leading to
integrands of the kind  $1/(\omega_1^2 \omega_2^2 \ldots \omega_n^2)$. Such a
factorization appears in the integrals resulting from the MSS, while the
integrands resulting from other topologies normally have a more complicated
structure. For example, the leading football two-pion-exchange diagrams
involving two Weinberg-Tomozawa vertices yield 
$1/(\omega_1 \omega_2 (\omega_1 + \omega_2 ))$. We can still use the same
machinery to obtain the potential in $r$-space employing the integral
representation 
\beq
\frac{1}{\omega_1 \omega_2 (\omega_1 + \omega_2)} = \frac{2}{\pi}
\int_0^\infty
d \beta \, \frac{1}{[\omega_1^2 + \beta^2] [\omega_2^2 + \beta^2]},
\eeq  
on the cost of introducing an additional power of $\pi$ in the denominator. 
But, then this power of $\pi$ means that ultimately this graph is {\it not} enhanced
compared to its $\chi$PT estimate. 
Similar arguments apply to the leading triangle and box diagrams
contributing to the two-pion exchange potential at NLO: neither of them is enhanced
by a power of $\pi$. In some cases, however, a sum of several 
non-MSS-type time-ordered graphs may lead to factorized expressions. The
simplest example is given by subleading two-pion exchange from the
triangle diagrams $\propto c_i$. In addition to the diagram contributing to the
MSS $\propto 1/(\omega_1^2 \omega_2^2)$, the remaining two time-ordered graphs
also yield
\beq
\frac{1}{\omega_1^2 \omega_2 (\omega_1 + \omega_2)} + \frac{1}{\omega_1
  \omega_2^2 (\omega_1 + \omega_2)} 
=  \frac{1}{\omega_1^2 \omega_2^2}.
\eeq
This is the reason why the result in
  Eq.~\eqref{simple_case} for $n=2$ actually gives  half the chiral NNLO potential.
It remains to be clarified whether similar patterns emerge at higher orders in
the loop expansion, and therefore whether there are other contributions which are enhanced in a similar fashion 
to those discussed here. 

Indeed, at each order in the chiral
expansion, there are many other topologies and contributions that have not been considered here. Again, one can argue that for non-MSS diagrams such as e.g.~the correlated
two-pion exchange, one generally does not expect enhancement by powers of
$\pi$, but this is not a proof. Furthermore, within the class of MSS time-ordered diagrams driven by the
$c_i$'s we did not take into account the energy dependence of the $c_i$-vertex
and have not considered the contributions driven by $c_4$. Although all of these points will change the quantitative results for the $NN$ potential, none of them should affect the qualitative conclusion. Poles at $r \sim 0.8\,{\rm fm}$ can still be expected in the full result, thanks to terms which are exponentially suppressed for $r > 1\,{\rm fm}$, but become comparable to the dominant two-pion-exchange parts of the $NN$ potential as the inter-nucleon distance decreases further.
\end{sloppypar}

\section{Conclusion}
\label{sec:conc}

The multiple-scattering series in its resummed form has been known and used for
decades. In this work we have looked at that result from an EFT point of view, examining
the divergence structure of the individual terms of
the series in detail. This is of particular
importance since a theoretical analysis of modern, high-accuracy, meson--nucleon
experiments calls for solid, quantitative understanding of the corresponding amplitudes.
The enhancement of MSS terms by powers of $\pi$ compared to their expected $\chi$PT size might lead to concerns about a related enhancement of the associated counterterms.

We have demonstrated analytically that, although the EFT integrals for
the MSS in momentum space are singular starting from the fourth term, under
certain circumstances all UV divergences cancel upon resummation, provided
that resummation is executed with a finite regulator in place.  The error
induced upon taking the regulator to infinity in the end is less than the size of the counterterm that
absorbs the dominant short-distance effects in the $\pi NN$ system---as must
be the case if the power counting for that counterterm is valid.  Therefore
--- at least in the case of a natural ($|a|\sim 1/\Lambda$) or a positive  scattering length,
e.g.  in the $\pi d$ case--- no enhanced counterterms are necessary.
Counterterms of normal $\chi$PT size will still complement the MSS
diagrams---as will other $\chi$PT graphs which are not of the MSS topology.
But, the $r$-space MSS expressions derived a long time ago based on
the Fourier transform of the integral equations in momentum space are
justified from an EFT point of view as the well-defined sum of a particular
class of graphs.

\begin{sloppypar}
Meanwhile, straightforward evaluation of the MSS in the scattering-length
approximation for the meson--nucleon $t$-matrix implies the appearance of an unphysical pole
in the $r$-space result (cut in the momentum-space result) in the case $a <
0$. For a natural scattering length this pole is outside the range of
applicability of the theory. However, for an unnaturally large scattering
length ($|a| \gg 1/\Lambda$), the scattering-length approximation for a
meson--nucleon $T$-matrix is not justified, and the inclusion of the unitarity, recoil,
and/or range corrections becomes necessary.  
Once this is done the pole may be shifted  towards the origin, and so move into a region 
outside the domain of applicability of the EFT, or it may even disappear completely.
However, additional investigations are necessary in order to confirm this conjecture.

This scenario might be expected to lead to difficulties for kaon--nucleus scattering, since $a_{KN}$ 
is negative and large enough in certain channels that the MSS may have a pole in the physical region. 
However, for an 
isovector-dominated meson--nucleon amplitude a pole never appears in the resummed series. In addition, 
in non-relativistic EFT, which is   used for $Kd$ scattering, 
the pole could appear only quite close to  the origin since $\Lambda|a|$ is only slightly larger than $1$. 
Therefore no enhanced counterterms are necessary in order to render the MSS for  kaon--nucleus 
 scattering sensible. 
\end{sloppypar}

Similar resummations of the MSS graphs which appear in the $NN$ potential are
also possible. In that case the potential {\it does} develop a pole at a
finite radius $r$. This implies that the $\chi$PT expansion for $V_{NN}$ has already broken down once
the pole appears. This suggests a limitation of $r > 1\,{\rm fm}$ for the
successful application of $\chi$PT to $V_{NN}(r)$, although this limit could be
lowered if dynamical Delta degrees of freedom are included in the EFT.

\section*{Acknowledgments}

\begin{sloppypar}
This research was supported by the DFG (SFB/TR 16, ``Subnuclear Structure of Matter''), DFG-RFBR grant (436 RUS 113/991/0-1),  
the Bonn-Cologne Graduate School of Physics and Astronomy, the DAAD,
the project ``Study of Strongly Interacting Matter''
(HadronPhysics3) under the 7th Framework Programme of the EU, the European Research Council
(ERC-2010-StG 259218 NuclearEFT), and 
the US Department of Energy (Office of Nuclear Physics, under contract No.~DE-FG02-93ER40756 with Ohio University). 
\end{sloppypar}

\end{document}